\documentclass[prl,twocolumn,amsmath,amssymb,superscriptaddress,showpacs]{revtex4}

\usepackage{graphicx}
\usepackage{epsfig}
\usepackage{epstopdf}

\begin{document}

\title{Multiband Transport in Bilayer Graphene at High Carrier Densities}
\author{Dmitri K. Efetov}
\author{Patrick Maher}
\author{Simas Glinskis}
\author{Philip Kim}
\affiliation{Department of Physics, Columbia University New York, NY 10027}
\date{\today}

\begin{abstract}
We report a multiband transport study of bilayer graphene at high carrier densities. Employing a poly(ethylene)oxide-CsClO$_4$ solid polymer electrolyte gate we demonstrate the filling of the high energy subbands in bilayer graphene samples at carrier densities $|n|\geq2.4\times 10^{13}$~cm$^{-2}$. We observe a sudden increase of resistance and the onset of a second family of Shubnikov de Haas (SdH) oscillations as these high energy subbands are populated. From simultaneous Hall and magnetoresistance measurements together with SdH oscillations in the multiband conduction regime, we deduce the carrier densities and mobilities for the higher energy bands separately and find the mobilities to be at least a factor of two higher than those in the low energy bands.

\end{abstract}

\maketitle


Multiband transport is common for many complex metals where different types of carriers on different pieces of the Fermi Surface (FS) carry electrical currents. Conduction in this regime is controlled by the properties of the individual subbands, each of which can have distinct mobilities, band masses, and carrier densities. Other changes to the single-band conduction model include inter-band scattering processes and mutual electrostatic screening of carriers in different subbands, which alters the effective strength of the Coulomb potential and hence adjusts the strength of electron-electron and electron-charged impurity interactions.

To understand electronic conduction in this regime, it is desirable to study the properties of the individual bands separately and compare these to the properties in the multiband regime. This was achieved in 2-dimensional electron gases (2DEGs) formed in GaAs quantum wells~\cite{ GaAs82}, where the subbands can be continuously populated and depopulated by inducing parallel magnetic fields. In these 2DEGs, an increased overall scattering rate due to interband scattering was observed upon the single- to multiband transition, \cite{GaAs90,GaAs97}, along with changes in the effective Coulomb potential which led to the observation of new filling factors in the fractional quantum Hall effect~\cite{Shayegan10}.

\begin{figure}[tbp]
\includegraphics[width=0.9\linewidth]{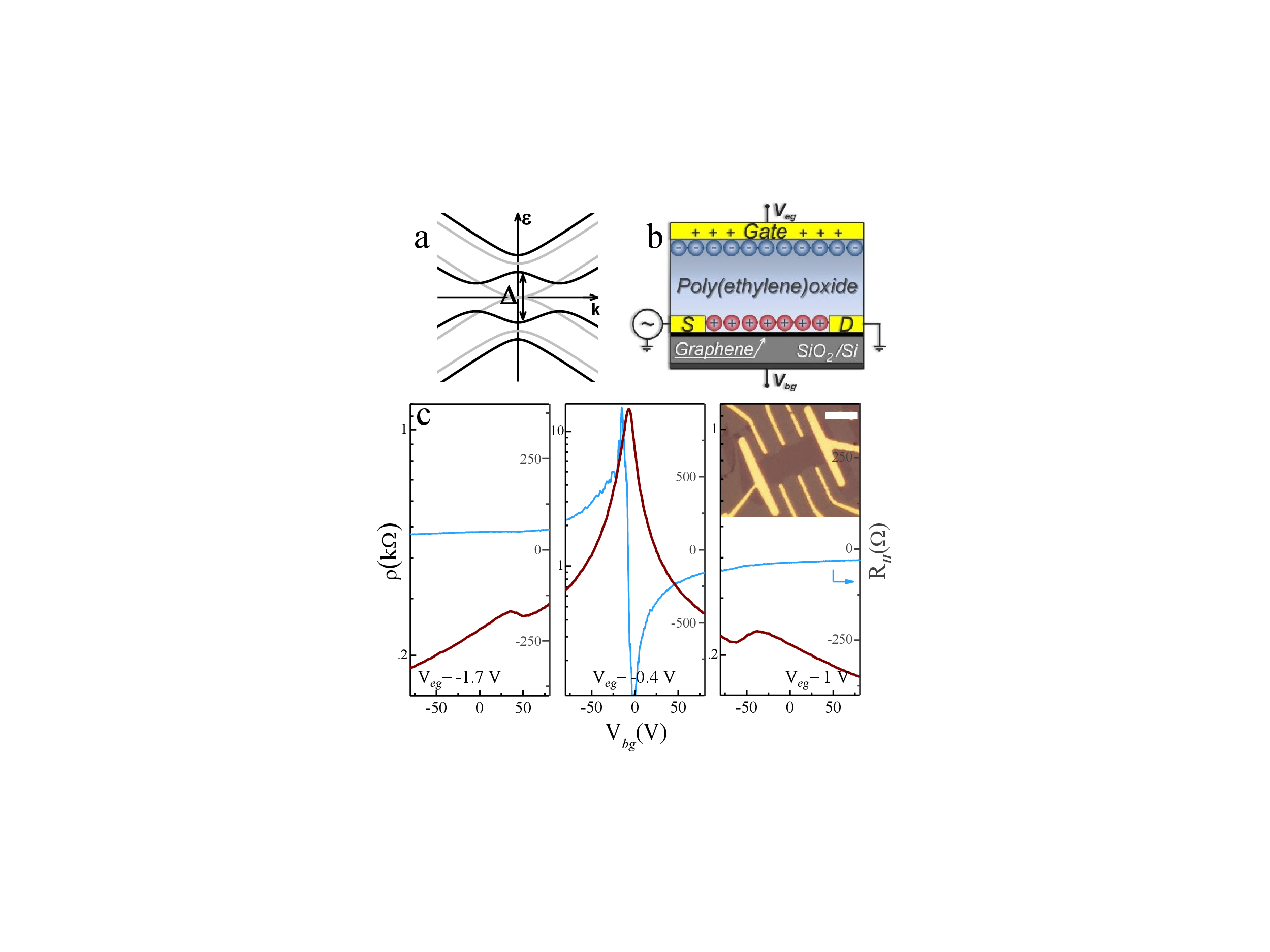}
\caption{(a) The tight-binding band structure of bilayer graphene for interlayer asymmetries $\Delta=0$ eV (gray) and $\Delta=0.6$ eV (black). (b) Schematic view of the double gated device, consisting of the SiO$_2$/Si back gate and the electrolytic top gate. Debye layers of Cs$^+$ or ClO$_4^-$ ions are formed $d\sim 1$~nm above the bilayer and the gate electrode, respectively. (c) Longitudinal resistivity and Hall resistance of the bilayer graphene device at $T=2$~K as a function of back gate voltage $V_{bg}$ for 3 different fixed electrolyte gate voltages $V_{eg}=$ -1.7, -0.4, and 1~V from left to right, corresponding to predoping levels of $n_{H}=$ (-2.9, 0, 2.9)$\times 10^{13}$~cm$^{-2}$. Inset shows an optical microscope image of a typical Hall Bar device (the scale bar corresponds to 5~$\mu$m).}
\label{Fig.1}
\end{figure}

\begin{figure*}[tbp]
\includegraphics[width=1.0\linewidth]{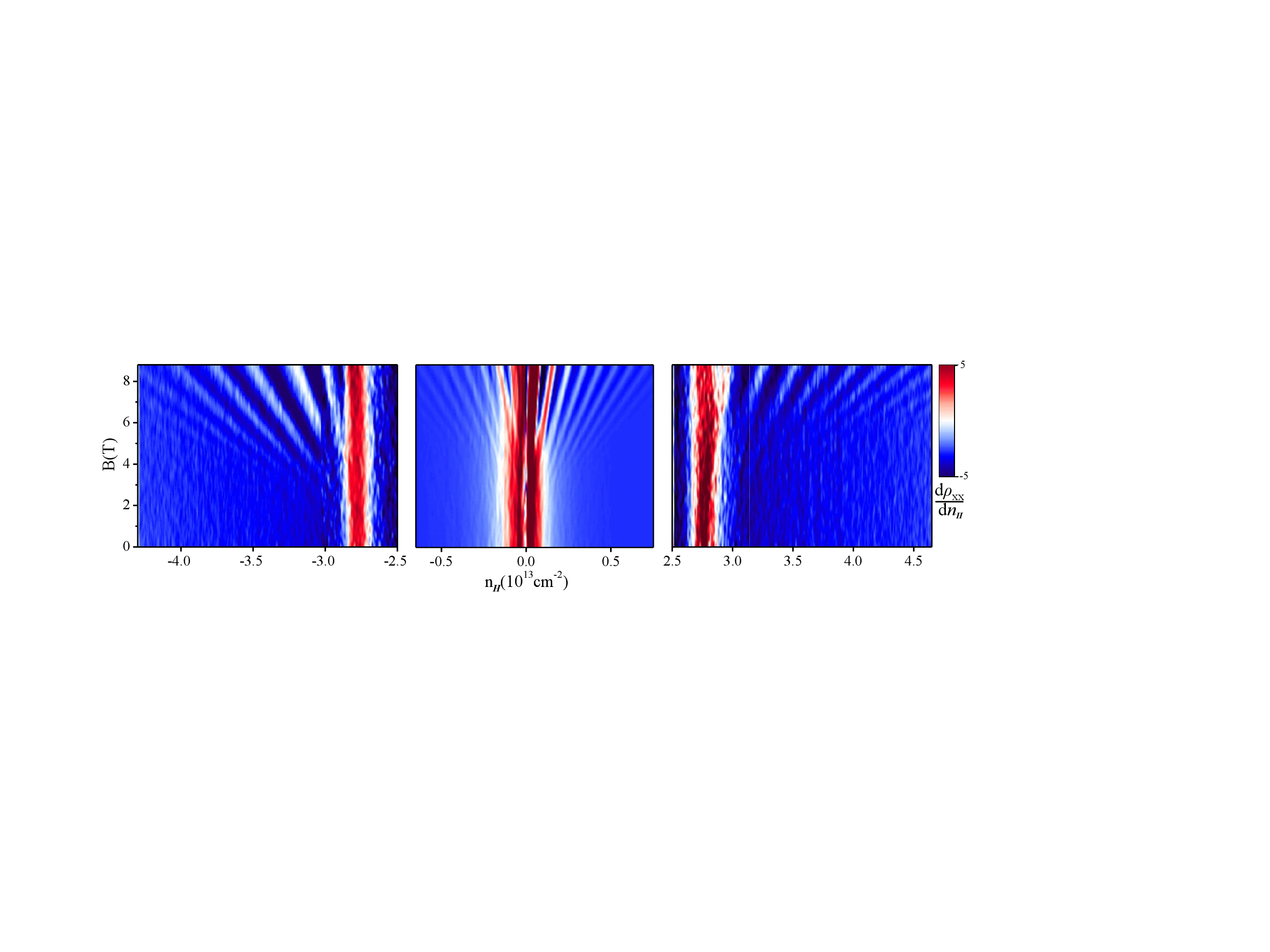}
\caption{Landau fan diagram of the differential longitudinal resistivity $\mathrm{d}\rho_{xx}/\mathrm{d}n_H$ for 3 different density ranges at $T=2$~K as a function of the Hall density $n_{H}$ and the magnetic field. (center) The SdH oscillations in the LES converge at the CNP and flatten out at higher $n_{H}$ due to decreasing LL separation. (left and right) For $|n_{H}|>2.6\times10^{13}$ cm$^{-2}$ additional SdH oscillations appear, originating at the resistivity spikes (red regions) that mark the onset of the HES.}
\label{Fig.2}
\end{figure*}

Bilayer graphene (BLG)\cite{Geim07,Geim08,McCann06,Rotenberg06,Novoselov07}, with its multiband structure and strong electrostatic tunability, offers a unique model system to investigate multiple band transport phenomena. BLG's four-atom unit cell yields a band structure described by a pair of low energy subbands (LESs) touching at the charge neutrality point (CNP) and a pair of high energy subbands (HESs) whose onset is $\sim\pm$0.4~eV away from the CNP (Fig.1(a)). Specifically, the tight binding model yields the energy dispersion \cite{McCann06}:
\begin{equation}
\epsilon_{1,2}^{\pm}(k)=\pm\sqrt{\frac{\gamma_{1}^{2}}{2}+\frac{\Delta^{2}}{4}+v_{F}^{2}{k}^{2}\pm\sqrt{\frac{\gamma_{1}^{4}}{4}+v_{F}^{2}{k}^{2}(\gamma_{1}^{2}+\Delta^{2})}}, \label{a1}
\end{equation}
where the upper and lower index indicates the conduction (+) and valence (-); and LES (1) and HES (2), $k$ is the wave vector measured from the Brillouin zone corner, $v_{F}\approx $10$^6$~m/s is the Fermi velocity in single layer graphene, $\gamma_{1}\approx0.4$~eV is the interlayer binding energy, and $\Delta$ is the interlayer potential asymmetry. Interestingly, since a perpendicular electric field $E$ across the sample gives rise to an interlayer potential difference $\Delta$, it opens up a gap in the spectrum of the LES~\cite{Rotenberg06,band gap,Fai09} and is furthermore predicted to adjust the onset energy of the HES. Whereas the LESs have been widely studied, the HESs, with their expected onset density of $n \gtrsim 2.4\times10^{13}$~cm$^{-2}$~\cite{Rotenberg06}, have thus far not been accessed in transport experiments. This can mainly be attributed to the carrier density limitations set by the dielectric breakdown of the conventional SiO$_2$/Si back gates, which do not permit the tuning of carrier densities above $n\approx0.7\times10^{13}$~cm$^{-2}$ ($\epsilon_F \approx$ 0.2~eV).

In this letter, we report multiband transport in bilayer graphene. Using an electrolytic gate, we were able to populate the HES of bilayer graphene, allowing for both the LES and HES to be occupied simultaneously. The onset of these subbands is marked by an abrupt increase of the sample resistivity, most likely due to the opening of an interband scattering channel, along with the appearance of a new family of Shubnikov-de Haas (SdH) oscillations associated with the HES. A detailed analysis of the magneto- and Hall resistivities in combination with the HES SdH oscillations in this regime enables us to estimate the carrier mobilities in each subband separately, where we observe a two-fold enhanced mobility of the HES carriers as compared to the LES carriers at the same band densities.

Bilayer graphene devices were fabricated by mechanical exfoliation of Kish graphite onto 300 nm thick SiO$_2$ substrates, which are backed by degenerately doped Si to form a back gate. The samples were etched into a Hall bar shape with a typical channel size of $\sim$ 5 $\mu$m and then contacted with Cr/Au (0.5/30 nm) electrodes through beam lithography (Fig. 1(c) inset). In order to access the HES we utilized a recently developed solid polymer electrolyte gating technique\cite{electrolyte,Iwasa08,Fai09,me10}, which was recently shown to induce carrier densities beyond values of $n>10^{14}$~cm$^{-2}$~\cite{me10} in single layer graphene. The working principle of the solid polymer electrolyte gate is shown in Fig.1(b). Cs$^+$ and ClO$_{4}^{-}$ ions are mobile in the solid matrix formed by the polymer poly(ethylene)oxide (PEO). Upon applying a gate voltage $V_{eg}$ to the electrolyte gate electrode, the ions form a thin Debye layer a distance $d\sim1$~nm away from the graphene surface. The proximity of these layers to the graphene surface results in huge capacitances per unit area $C_{eg}$, enabling extremely high carrier densities in the samples. While CsClO$_{4}$ has almost the same properties as the typically used LiClO$_{4}$ salt, we find a reduced sample degradation upon application of the electrolyte on top of the sample, resulting in considerably higher sample mobilities.

One major drawback of the electrolyte gate for low temperature studies is that it cannot be tuned below $T<250$~K, where the ions start to freeze out in the polymer and become immobile (though leaving the Debye layers on the bilayer surface intact)~\cite{Iwasa08,me10}. A detailed study of the density dependent transport properties at low temperatures can therefore be quite challenging. In order to overcome this issue, we employ the electrolyte gate just to coarsely tune the density to high values ($|n|< $10$^{14}$ cm$^{-2}$) at $T=300$~K, followed by an immediate cooldown to $T=2$~K. We then use the standard SiO$_2$/Si back gate to map out the detailed density dependence of the longitudinal sheet resistivity $\rho_{xx}$ and the Hall resistance $R_{H}$, from which we extract the total carrier density of the sample $n_{H} =B/eR_{H}$, with $B$ the magnetic field and $e$ the electron charge. Here we find the back gate capacitance to be $C_{bg}=141$~aF/$\mu$m$^2$, almost unaltered by the presence of the Debye layers on top of the sample.

In this experiment, we have measured $\rho_{xx}$ and $R_{H}$ of more than 3 BLG devices as a function of the back gate voltage $V_{bg}$ at various fixed $V_{eg}$ corresponding to the wide density range of $n_{H}\sim \pm 8\times10^{13}$ cm$^{-2}$. Fig.1(c) shows $\rho_{xx}$ and $R_{H}$ for a representative device for 3 selected cool-downs at $V_{eg}=$ -1.7, -0.4, 1~V from left to right, corresponding to a pre-doping level of $n_{H} =$(-2.9, 0, 2.9)$\times 10^{13}$~cm$^{-2}$. For low doping levels ($V_{eg}=$-0.4~V, Fig.1(c) middle) we observe the expected Dirac Peak in $\rho_{xx}$ and the ambipolar transition of $R_{H}$ as $V_{bg}$ sweeps through the CNP. Away from the CNP, $\rho_{xx}$ and $R_{H}$ decrease as $|n_{H}|$ increases, as was observed before in BLG samples~\cite{Novoselov07}. For the strongly pre-doped gate sweeps however (Fig.1(c) left and right), we observe a rather unexpected non-monotonic feature in the sample resistivity. Instead of a monotonic decrease of $\rho_{xx}$ with increasing $|n_{H}|$, it exhibits an abrupt increase by $\sim10\%$ symmetrically at both electron and hole sides at $n^* \sim |n_{H}|= 2.6 \times 10^{13}$~cm$^{-2}$, a carrier density which is consistent with theoretical expectations for the onset density of the HES~\cite{McCann06,Rotenberg06}. A similarly increasing resistivity at the opening of a new subband was previously observed in 2D electron systems formed in wide GaAs quantum wells~\cite{GaAs82,GaAs90,GaAs97}, where it was attributed to a decreased overall scattering time $\tau$ due to the opening of an additional inter-band scattering channel as the new subbands are populated. Such an inter-band scattering mechanism between the LES and the HES is also expected to give rise to a resistivity increase upon filling of the HES in BLG samples, making it a likely candidate for the origin of the observed resistivity increase. However, considering the strong differences between the 2D electron gases in GaAs quantum wells and in BLG, including vastly different densities of states, mobilities, and electron energies, more theoretical work needs to be done to conclusively determine the origin of this resistivity increase.


The electronic structure of the LES and HES can be further investigated by studying the effect of the magnetic field $B$ on the longitudinal resistivity $\rho_{xx}(B)$ in the various density ranges. Fig.2 shows the Landau fan diagram of the differential sheet resistivity $\mathrm{d}\rho_{xx}/\mathrm{d}n_{H}$ as a function of $B$ and $n_{H}$. Close to the CNP (Fig.2 center) the SdH oscillations in the two LES are quite pronounced, but with increasing density their amplitude quickly decays as the energy separation of the Landau Levels (LL) decreases. Above the onset of the HES (Fig.2 left and right), marked by the ``spikes" of increased resistivity (here the red regions) however, we observe another set of SdH oscillations which form LL fans converging into the onset point of the HES.


In order to analyze the SdH oscillations, we now plot the $\rho_{xx}(B)$ traces for various fixed $n_{H}$ as a function of the inverse magnetic field $B^{-1}$. Fig. 3(a) displays three exemplary traces at different $n_H$ above the onset density of the HES. All traces show periodic oscillations in $B^{-1}$ allowing us to obtain the SdH density, $n_{SdH}=\frac{4e}{h}\Delta(B^{-1})$, assuming that each LL is both spin and valley degenerate. Whereas for all $|n_H|<n^*$ we find that the obtained $n_{SdH} \approx n_{H}$, indicating that the SdH oscillations are solely from a single band (i.e., the LES), for $|n_H|>n^*$ the obtained $n_{SdH}$ values are much smaller than the simultaneously measured $n_{H}$ values. This behavior can be well explained by assuming that these SdH oscillations reflect only the small fraction of charge carriers lying in the HES. For $|n_H|>n^*$ we hence are able to extract the occupation densities of the LES ($n_{LES}$) and HES ($n_{HES}$) from $n_{LES}=n_H-n_{SdH}$ and $n_{HES}=n_{SdH}$. Fig. 3(b) shows the $|n_{LES}|$ and $|n_{HES}|$ in this regime as a function of the total carrier density $|n_H|$. For each fixed $V_{eg}$, the obtained $|n_{LES}|$ and $|n_{HES}|$ increase as $|n_{H}|$ increases (adjusted by $V_{bg}$), for both electrons and holes. Interestingly, we notice that the $|n_{LES}(n_H)|$ are slightly larger for larger $|V_{eg}|$ while the trend is opposite for the HES, i.e. $|n_{HES}(n_H)|$ are smaller for larger $|V_{eg}|$, even though their $n_H$ values are in similar ranges. These general trends can be explained by an increase of the interlayer potential difference $\Delta$ for increased values of $|V_{eg}|$, which are predicted by the tight-binding model in Eq.1 to result in an increase of the onset density (energy) of the HES.

\begin{figure}[tbp]
\includegraphics[width=1.0\linewidth]{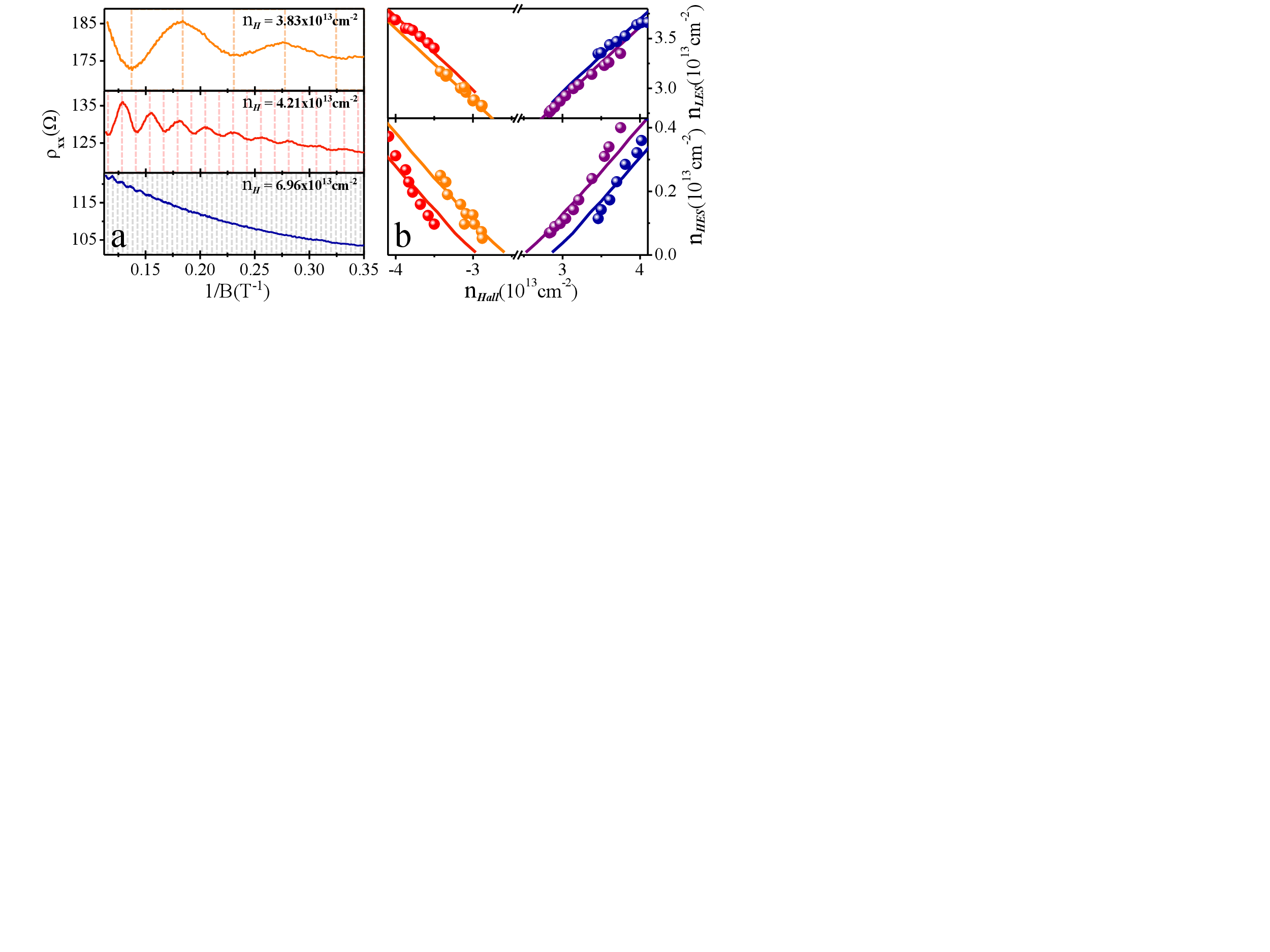}
\caption{(a) Exemplary traces of the longitudinal resistivity as a function of inverse magnetic field at fixed values of $n_{H}$ beyond the onset of the HES. (b) Carrier densities inferred from the SdH oscillations vs. the overall Hall densities $n_{H}$, from 4 cooldowns at different set electrolyte gate voltages $V_{eg}=$ -2~V (red), -1.7~V (orange), 1~V (purple), 1.4~V (blue). (bottom) $|n_{HES}|$ vs. $n_{H}$, fitted with theoretical expectations for the HES. (top) $|n_{LES}|$ vs. $n_{H}$, fitted with theoretical expectations for the LES. Line traces correspond to theoretical fits for different values of $\Delta=$ 0.31~eV (red), 0.17~eV (orange), 0.13~eV (purple), 0.26~eV (blue).}\label{Fig.3}
\end{figure}

While a precise quantitative determination of the expected shift in the onset density of the HES as a function of $V_{eg}$ and $V_{bg}$ requires a self-consistent calculation of $\Delta(V_{eg}, V_{bg})$ and would go beyond the scope of this paper, we can still qualitatively test the above prediction. This is possible since $\Delta$ is mostly controlled by $V_{eg}$, which has a much stronger coupling to the BLG sample than the $V_{bg}$, thus allowing us to approximately treat $\Delta$ as a constant for fixed $V_{eg}$. Since the experimental traces displayed in Fig.3(b) correspond to different values of $V_{eg}$ but the same ranges of $V_{bg}$, $\Delta$ is different for each trace and can be extracted from the theoretical fits from Eq.~1, with $\Delta$ as the only fitting parameter. Indeed for all 4 traces we find good agreement with the theoretical fits; we clearly observe an enhanced onset density (energy) for the traces with larger set potential differences across the sample, which is in good qualitative agreement with theoretical predictions.


We now turn our attention to the transport properties of BLG in the limit of $n_H > n^*$. The filling of these sub-bands creates a parallel transport channel in addition to the one in the LES, thus defining the transport properties in this regime by two types of carriers with distinct mobilities $\mu_{1,2}$, effective masses $m^{*}_{1,2}$ and subband densities $n_{1,2}$ (here the index corresponds to the LES(1) and HES(2))~\cite{Fuhrer08,vanHouten88}. In sharp contrast to a single band Drude model, where $\rho_{xx}(B)$ does not depend on the $B$ field, in a two-carrier Drude theory it is expected to become strongly modified, resulting in a pronounced $B$ field dependence \cite{Ashcroft76}:
\begin{equation}
\rho_{xx}(B)=\frac{n_1\mu_1+n_2\mu_2+(n_1\mu_1\mu_2^2+n_2\mu_2\mu_1^2)B^2}{e((n_1\mu_1+n_2\mu_2)^2+\mu_1^2\mu_2^2(n_1+n_2)^2B^2)}, \label{a2}
\end{equation}

Fig.4(a) shows magnetoresistance traces for different fixed Hall densities $n_{H}$. Close to the CNP, where only the LES are populated (Fig.4(a) black trace), the $\rho_{xx}(B)$ traces are nearly flat as expected from the one-fluid Drude theory. When the density is increased and the HES starts to fill up, however, we observe a smooth transition to an approximately parabolic $B$ field dependence, resulting in a strong increase of $\rho_{xx}$ of up to 25$\%$ from 0~T to 8~T. Using the previously extracted carrier densities in the two bands $n_{1,2}$ we can now fit the $\rho_{xx}(B)$ traces with the two-carrier Drude model in Eq.2, with the mobilities of the two subbands $\mu_{1,2}$ as the only fitting parameters. As shown in Fig.4(b) the experimental finding are in excellent agreement with the theory, allowing us to deduce the values of $\mu_{1,2}$ with good accuracy. Moreover, the ability to extract the mobilities of the HES allows us now to characterize the HES in more detail.

In Fig.4(c) we plot the extracted mobilities of the HES $\mu_{2}$ against the carrier density in the HES $n_{2}$ and compare it to the mobilities $\mu_{1}$ of the LES at a similar range of subband densities in the LES $n_{1}$. We find that the mobilities in the HES are at least a factor of two higher than those in the LES. Considering that the effective carrier masses are similar for the LES and the HES, this feature of the HES may be due to the enhanced screening of charged impurity scatterers at higher carrier densities, effectively reducing the scattering rate of the HES carriers on these scatterers. A more detailed theoretical study is required, however, to undertake a quantitative analysis of this problem.


\begin{figure}[tbp]
\includegraphics[width=0.9\linewidth]{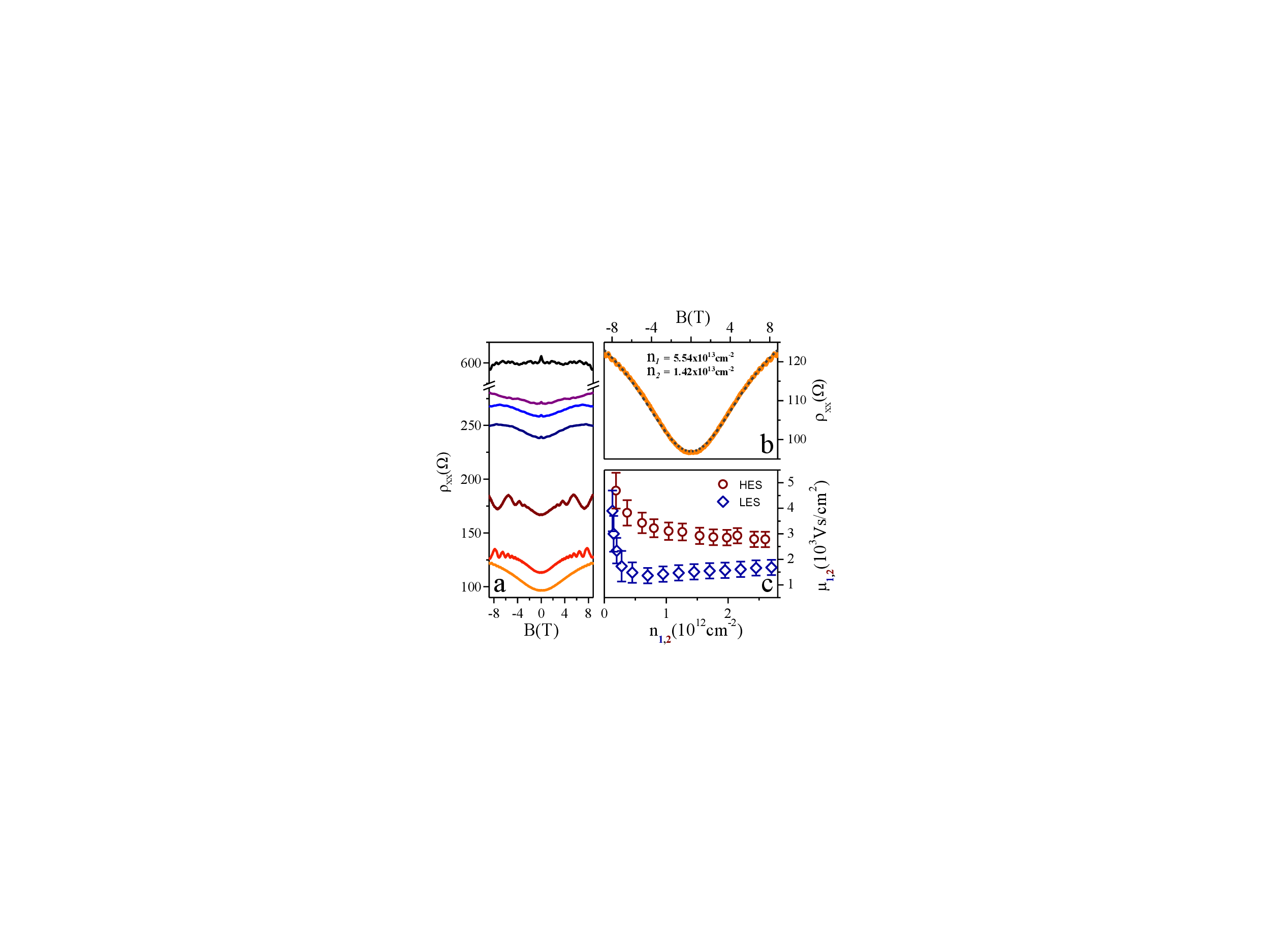}
\caption{(a) Longitudinal resistivity $\rho_{xx}(B)$ as a function of magnetic field for $n_{H}=$ (0.84, 2.88, 3.00, 3.31, 3.83, 4.21, 6.96)$\times10^{13}$~cm$^{-2}$, from top to bottom. The $\rho_{xx}(B)$ traces undergo a smooth transition from a nearly $B$ independent behavior when the HES is empty (below $n_{H}<n^*\sim2.6\times10^{13}$ cm$^{-2}$), to a strong, non-trivial $B$ dependence when the HES is occupied. (b) An exemplary $\rho_{xx}(B)$ trace at $n_{H} = $6.96$\times 10^{13}$~cm$^{-2}$ and $n_{SdH}=$1.42$\times 10^{13}$~cm$^{-2}$ with accompanying fit (dashed line) from Eq.2, using the mobilities in the LES $\mu_{1}=541$~Vs/cm$^{2}$ and the HES $\mu_{2}=2428$~Vs/cm$^{2}$ as fitting parameters. (c) The mobilities $\mu_{1,2}(n)$ as extracted from $\rho_{xx}(B)$ traces at various fixed $n_{H}$ as a function of the density in the individual subbands.}
\label{Fig.4}
\end{figure}

In conclusion, using a polymer electrolyte gate we have achieved two-band conduction in bilayer graphene. We have found that the filling of these bands above a Hall density of $|n_{H}|>2.4\times 10^{13}$~cm$^{-2}$ is marked by an increase of the sample resistivity by $\sim10\%$ along with the onset of SdH oscillations. From simultaneous Hall and magnetoresistivity measurements, as well as the analysis of the SdH oscillations in the two carrier conduction regime, we have characterized the distinct carrier densities and mobilities of the individual subbands, where we have found a strongly enhanced carrier mobility in the HES of bilayer graphene.

The authors thank I.L. Aleiner, E. Hwang and K.F. Mak for helpful
discussion. This work is supported by the AFOSR MURI, FENA, and DARPA
CERA. Sample preparation was supported by the DOE
(DE-FG02-05ER46215).


\begin{thebibliography}{99}

\bibitem{GaAs82} H. L. Stormer et. al., Sol. St. Com., \textbf{41}, 707-709 (1982).

\bibitem{GaAs97} G. R. Facer et al., Phys. Rev. B \textbf{56}, 10036-10039 (1997).

\bibitem{GaAs90} D. R. Leadley et al., Semi. Sci. Tech. \textbf{5}, 1081-1087 (1990).

\bibitem{Shayegan10} J. Shabani et al., arXiv:1004.0979v2 (2010).

\bibitem{Geim07} A.~K. Geim and K.~S. Novoselov, Nat Mater \textbf{6}, 183
(2007).

\bibitem{Geim08} A.~K. Geim and P. Kim, Scientific American \textbf{298}, 68
(2008).

\bibitem{McCann06} E. McCann et. al., Phys. Rev. B \textbf{74}, 161403 (2006).

\bibitem{Rotenberg06} T. Ohta, et.al., Science \textbf{313}, 951 (2006).

\bibitem{Novoselov07} K. S. Novoselov, et. al., Nature Physics \textbf{2}, 177 - 180 (2006)
(2007).

\bibitem{band gap} E. V. Castro et al., Phys. Rev. Lett. \textbf{99}, 216802 (2007); J. B. Oostinga et al., Nature Mater. \textbf{7}, 151 (2008); Y. Zhang et al., Nature (London) \textbf{459}, 820 (2009);

\bibitem{electrolyte} Matthew J. Panzer et.al., Advanced Materials
\textbf{20}, 3177 - 3180 (2008);  A. Das et.al., Nature Nanotechnology \textbf{3}, 210 - 215 (2008); J. Yan et. al., Phys. Rev. B \textbf{80}, 241417 (2009);

\bibitem{Iwasa08} K. Ueno et.al., Nature Materials \textbf{7}, 855 - 858 (2008);

\bibitem{Fai09} Kin Fai Mak et.al., Phys. Rev. Lett. \textbf{102}, 256405 (2009);

\bibitem{me10} Dmitri K. Efetov and Philip Kim, Phys. Rev. Lett. \textbf{105}, 256805 (2010).

\bibitem{Fuhrer08} Sungjae Cho and Michael S. Fuhrer, Phys. Rev. B \textbf{77}, 084102 (2008).

\bibitem{vanHouten88} H. van Houten et.al., Phys. Rev. B \textbf{37}, 2756-2758 (1988).

\bibitem{Ashcroft76} Neil W. Ashcroft and N. David Mermin, Thomson Learning Inc. (1976).

\end{thebibliography}
\end{document}